\documentclass[12pt]{article}
\usepackage{graphicx}
\begin{document}

\begin{center}
{\Large \bf  Standing waves in the Lorentz-covariant world}

\vspace{3ex}

Y. S. Kim\footnote{electronic address: yskim@physics.umd.edu}\\
Department of Physics, University of Maryland,\\
College Park, Maryland 20742, U.S.A.\\

\vspace{3ex}

Marilyn E. Noz \footnote{electronic address: noz@nucmed.med.nyu.edu}\\
Department of Radiology, New York University,\\ New York, New York 10016, U.S.A.\\

\end{center}

\vspace{3ex}

\begin{abstract}

When Einstein formulated his special relativity, he developed his
dynamics for point particles.  Of course,
many valiant efforts have been made to extend his relativity to
rigid bodies, but this subject is forgotten in history.
This is largely because of the emergence of quantum mechanics with
wave-particle duality.  Instead of Lorentz-boosting rigid bodies,
we now boost waves and  have to deal with Lorentz transformations
of waves.  We now have some understanding of plane waves or
running waves in the covariant picture, but we do not yet have
a clear picture of standing waves.  In this report, we show that
there is one set of standing waves which can be Lorentz-transformed
while being consistent with all physical principle of quantum
mechanics and relativity.  It is possible to construct a
representation of the Poincar\'e group using harmonic oscillator
wave functions satisfying space-time boundary conditions.  This set
of wave functions is capable of explaining the quantum bound state
for both slow and fast hadrons.  In particular it can explain the
quark model for hadrons at rest, and Feynman's parton model hadrons
moving with a speed close to that of light.
\end{abstract}

\newpage

\section{Introduction}\label{intro}

Einstein formulated his special relativity one hundred years
ago while making Newtonian mechanics consistent with the
Lorentz-covariant world.  In so doing, he derived his
energy-momentum relation valid for both massive and massless
particles.  Einstein of course formulated his theory for point
particles.  Since then, there have been efforts to understand
special relativity for rigid particles with non-zero size,
without any tangible results.  On the other hand, the emergence
of quantum mechanics made the rigid-body problem largely
irrelevant.

Because of the wave-particle duality, quantum mechanics is
sometimes called wave mechanics.  Instead of rigid bodies,
we talk about wave packets and standing waves.  The issue
becomes whether those waves can be made Lorentz-covariant.

Of course, here, the starting point is the plane wave,
which can be written as
\begin{equation}
e^{ip \cdot x} = e^{i(\vec{p}\cdot\vec{x} - Et)} .
\end{equation}
Since it takes the same form for all Lorentz frames, we do not
need any extra effort to make it covariant.

Indeed, the S-matrix derivable from the present form of quantum
field theory calls for calculation of all S-matrix quantities
in terms of plane waves. Thus, the S-matrix is associated with
perturbation theory or Feynman diagrams.  Indeed, Feynman
propagators are written in terms of  plane waves on the mass
shell.

%------------------------------------------------------------------------
\begin{figure}[thb]
\centerline{\includegraphics[scale=0.7]{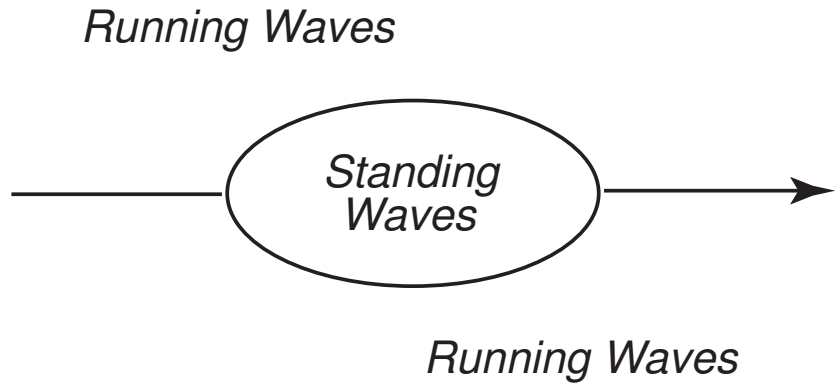}}
\vspace{5mm}
\caption{Running waves and standing waves in quantum theory.  If a
particle is allowed to travel from infinity to infinity, it corresponds
to a running wave according to the wave picture of quantum mechanics.
If, on the other hand, it is trapped in a localized region, we have
to use standing waves to interpret its location in terms of
probability distribution.}\label{dff11}
\end{figure}
%----------------------------------------------------------------------

We should realize however that the S-matrix formalism is strictly
for running waves, starting from a plane wave from one end of the
universe and ending with another plane wave at another end.  How
about standing waves?  This question is illustrated in
Fig.~\ref{dff11}.  Of course, standing waves can be regraded as
superpositions of running waves moving in opposite directions.
However, in order to guarantee localization of the standing waves, we
need a spectral function or boundary conditions.  The covariance of
standing waves necessarily involve the covariance of boundary
conditions or spectral functions.  How much do we know about this
problem?

The purpose of this paper is to examine this problem systematically.
When we talk about standing waves in quantum mechanics, we start
with two standard examples, namely harmonic oscillators and
particles bound by hard walls separated by a space-like distance.
For the hard walls, we do not know how to deal with the covariance
of the boundary conditions, and we are not able to report
anything in this paper.

For harmonic oscillators, boundary conditions are smooth, it might
be possible to impose a localization condition in a Lorentz-covariant
manner.  This possibility was considered by a number of great
physicists in the past, including Paul A. M. Dirac~\cite{dir45},
Hideki Yukawa~\cite{yuka53}, and Richard Feynman and his
colleagues~\cite{fkr71}.  The paper of Feynman {\it et al} was
written after Gell-Mann's formulation of the quark
model~\cite{gell64}, and is much closer to the real world.

Therefore, in this report, we start with the Lorentz-invariant
differential equation given by Feynman {\it et al.}~\cite{fkr71}.
Our first step is to make up mathematical deficiencies of this paper,
and to construct a set of covariant harmonic oscillator wave functions.
We then attach physical interpretations to these wave functions.
We point out that the covariant oscillator formalim satisfies all the
known rules of quantum mechanics and special relativity, as the
present form of quantum field theory does.

In addition, we point out that the covariant oscillator formalism can
explain the quark model for hadrons when they are at rest or slow, and
that the same formalism leads to Feynman's parton model when they
move with speed close to that of light.  Indeed, the quark model and
the parton model are two limiting cases of one covariant entity.

In Sec~\ref{feynm}, it is noted that there are running waves and
standing waves in quantum mechanics.  While it is easy to
Lorentz-boost running waves, it requires covariance of boundary
conditions to understand fully standing waves.  In Sec.~\ref{symm},
we discuss the space-time symmetry applicable to standing waves
in the Lorentz-covariant regime.  It is pointed out that this
symmetry is dictated by Wigner's little group~\cite{wig39,knp86}
for massive particles.

In Sec.~\ref{covham}, it is shown possible to construct a set of
harmonic oscillator wave functions, which can be Lorentz-boosted.
It is shown that these wave functions are compatible with all
known rules of quantum mechanics and special relativity.
As a physical application of this covariant harmonic oscillator
formalism, it is shown in Sec.~\ref{par} that the quark and parton
models are two different manifestation of the same covariant
entity.

\section{Scattering States and Bound States}\label{feynm}

We are now facing the problem of whether the basic concept of quantum
mechanics survives in Einstein's Lorentzian world.  By now, it is safe
to assume that Feynman diagrams serve our purpose well for scattering
states. Feynman diagrams are possible because the covariant form for
plane waves is quit trivial.

How about bound states?  In order to understand the bound-state problem,
we have to understand standing waves in the covariant world as indicated
in Fig.~\ref{dff11}.  In his talk presented at the 1970 April meeting of
the American physical society held in Washington, DC, Feynman stunned
the audience by saying that Feynman diagrams are not applicable
to bound state problems~\cite{feyn70,kim66}.  He suggested harmonic
oscillators for a possible solution.

%------------------------------------------------------------------------
\begin{figure}[thb]
\centerline{\includegraphics[scale=0.7]{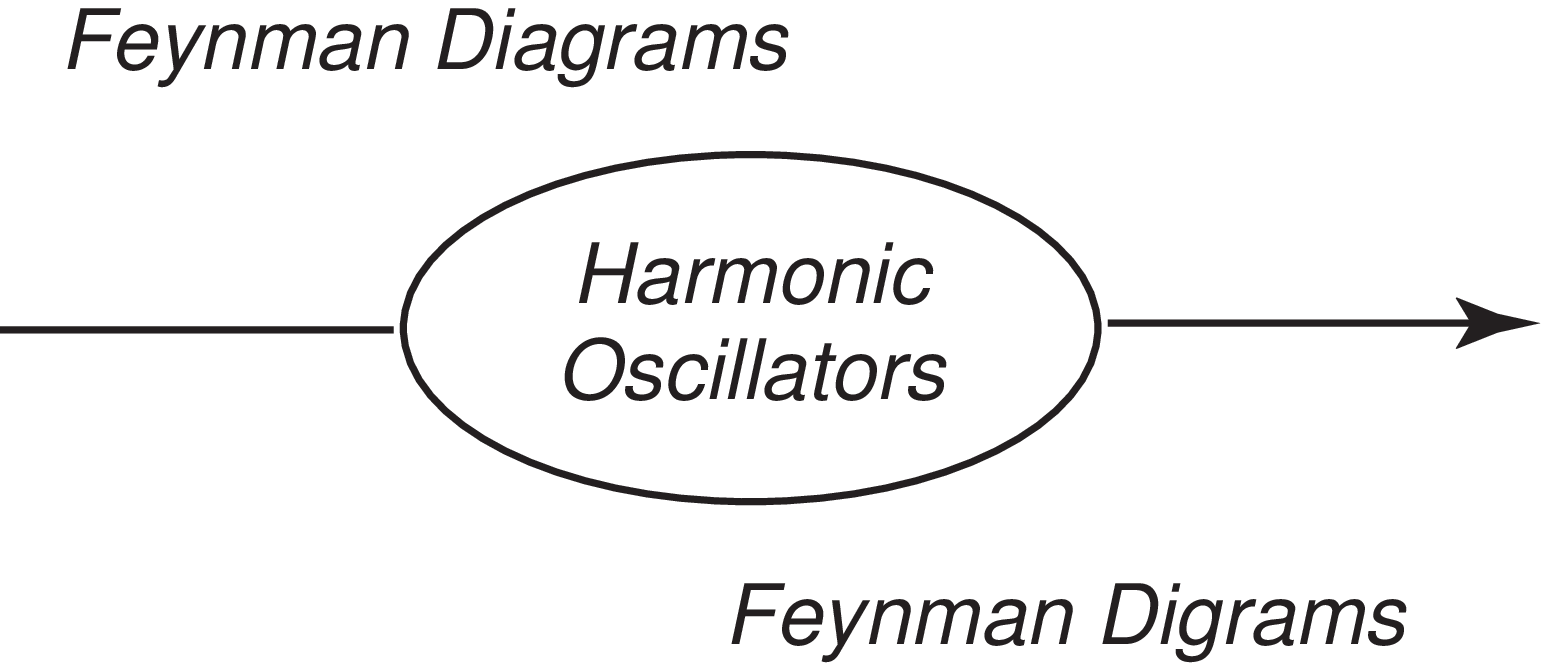}}
\vspace{5mm}
\caption{Feynman's roadmap for combining quantum mechanics with special
relativity.  Feynman diagrams work for running waves, and they provide
a satisfactory resolution for scattering states in Einstein's world.
For standing waves trapped inside an extended hadron, Feynman suggested
harmonic oscillators as the first step.}\label{dff33}
\end{figure}
%-------------------------------------------------------------------------

We can summarize what Feynman said in Fig.~\ref{dff33}.
Feynman's point was that, while plane-wave approximations in terms of
feynman diagrams work well for relativistic scattering problems, they
are not applicable to bound-state problems.  For bound-state problems,
we should perhaps try harmonic oscillator wave functions.
Feynman's 1970 talk was later published in the paper of Feynman, Kislinger,
and Ravndal in the Physical Review~\cite{fkr71}.

Although this paper contained the above mentioned original idea of Feynman,
it contains serious mathematical flaws.
Feynman {\it et al.} start with a Lorentz-invariant differential equation
for the harmonic oscillator for the quarks bound together inside a hadron.
For the two-quark system, they write the wave function of the form
\begin{equation}
\exp{\left\{{-1 \over 2}\left(z^2 - t^2 \right) \right\}} ,
\end{equation}
where $z$ and $t$ are the longitudinal and time-like separations between the
quarks.  This form is invariant under the boost, but is not normalizable in
the $t$ variable.

On the other hand, the Gaussian form
\begin{equation}
\exp{\left\{{-1 \over 2}\left(z^2 + t^2 \right) \right\}}
\end{equation}
also satisfies Feynman's Lorentz-invariant differential equation.
This Gaussian function is normalizable, but is not invariant under
the boost.  However, the word ``invariant'' is quite different
from the word ``covariant.''  The above form can be covariant
under Lorentz transformations.  We shall get back to this
problem in Sec.~\ref{covham}.

Feynman {\it et al.} studied in detail the degeneracy of the
three-dimensional harmonic oscillators, and compared with the
observed experimental data.  Their work is complete and thorough,
and is consistent with the $O(3)$-like symmetry dictated by
Wigner's little group for massive particles~\cite{wig39,knp86}.
Yet, Feynman {\it et al.} make an apology that the symmetry is
not $O(3,1)$.  This unnecessary apology causes a confusion not
only to the readers but also to the authors themselves, and makes
the paper difficult to read.

\section{Space-time Symmetry of Standing Waves}\label{symm}

As was noted in Sec.~\ref{feynm}, it is trivial to Lorentz-transform
plane waves.   How about superposition of plane waves?  We have to
deal with the Lorentz covariance of their spectral functions.  This
is not an easy problem for standing waves consisting of waves moving
in opposite directions.  We shall come back to this problem in
Sec.~\ref{covham}.  In this section, we shall study the space-time
symmetry of standing waves.

In the Lorentz-covariant world, the word standing wave means that
there is at least one Lorentz frame in which the amplitude is
non-zero in a localized spacial region, and this localization region
stays at the same place independent of time.  We shall call this
Lorentz frame ``the rest frame'' for the standing wave.
How would this standing wave look to an observer moving with a
constant velocity?  We can safely say the whole system will move
with a constant velocity in the opposite direction.  How about the
shape of the standing wave?  Is the concept of localization
preserved under Lorentz boosts?

\begin{figure}[thb]
\centerline{\includegraphics[scale=0.7]{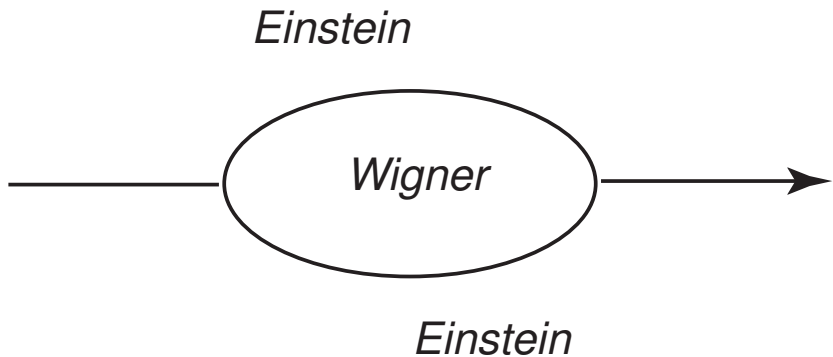}}
\vspace{5mm}
\caption{Wigner in Einstein's world.  Einstein formulates special
relativity whose energy-momentum relation is valid for point particles
as well as particles with internal space-time structure.  It was Wigner
who formulated the framework for internal space-time symmetries by introducing his
little groups whose transformations leave the four-momentum of a given particle
invariant.}\label{dff22}
\end{figure}

In order to tackle this problem, we have to understand the
space-time symmetry of this localized system.  The Lorentz group
applicable to a free particle has six parameters
corresponding three rotations around and three boosts along the
three orthogonal spatial directions.  Once the system is given in
a specific Lorentz frame, the system has only three degrees of
freedom, as is seen in Wigner's 1939 paper on his little
groups~\cite{wig39}.   The system regains all of the six degrees
of freedom when we add the three degrees of freedom to boost in
three independent directions~\cite{kim61b}.  Indeed, the bound
state or the standing wave has only three rotational space-time
degrees of freedom in the Lorentz frame in which it is at rest.
This picture of space-time symmetry is illustrated in Fig.~\ref{dff22}

%------------------------------------------------------------------------

%----------------------------------------------------------------------

This aspect of Wigner's little group has already been studied in
the past.  Let us see where the present problem stands in the
development of this subject.

Since Einstein introduced the Lorentz covariant space-time symmetry,
his energy momentum relation $E = \sqrt{p^2 + m^2}$ has been proven
to be valid for not only point particles, but also particles with
internal space-time structure defined by quantum mechanics.  Particles
can have quantized spins if they are at rest or they are slowly moving.
If, on the other hand, the particle is massless and moves with speed
of light, it has its helicity which is the spin parallel to its
momentum and gauge degree of freedom.

%-------------------------------------------------------------------
\begin{table}[thb]
\caption{Massive and massless particles in one package.  Wigner's
little group unifies the internal space-time symmetries for massive and
massless particles.  It is a great challenge for us to find
another unification: the unification of the quark and parton pictures in
high-energy physics.}\label{einwig}
\vspace{3mm}
 \begin{center}
\begin{tabular}{lccc}
\hline
{}&{}&{}&{}\\
{} & Massive, Slow \hspace{6mm} & COVARIANCE \hspace{6mm}&
Massless, Fast \\[4mm]\hline
{}&{}&{}&{}\\
Energy- & {}  & Einstein's & {} \\
Momentum & $E = p^{2}/2m$ & $ E = [p^{2} + m^{2}]^{1/2}$ & $E = p$
\\[4mm]\hline
{}&{}&{}&{}\\
Internal & $S_{3}$ & {}  &  $S_{3}$ \\[-1mm]
Space-time &{} & Wigner's  & {} \\ [-1mm]
Symmetry & $S_{1}, S_{2}$ & Little Group & Gauge Trans. \\[4mm]\hline
{}&{}&{}&{}\\
Relativistic & {} & One  &  {} \\[-1mm]
Extended & Quark Model & Covariant  & Parton Model\\ [-1mm]
Particles & {} & Theory &{} {} \\[4mm]\hline

\end{tabular}

\end{center}
\end{table}
%------------------------------------------------------------------------

Table~\ref{einwig} summarizes the covariant picture of the present
particle world.  The second row of this table indicates that the
spin symmetry of slow particles and the helicity-gauge symmetry of
massless particles are two limiting cases of one covariant entity
called Wigner's little group.  This issue has been extensively
discussed in the literature~\cite{kiwi90jm}.

Let us then concentrate on the third row of Table~\ref{einwig}.
After Einstein formulated his special relativity, a pressing problem
was to see whether his relativistic dynamics can be extended to
rigid bodies as in the case of Newton's sun and earth and their
rotations.  Their rotations are translated into particle spins in
quantum mechanics.  Their sizes can be translated into the
width of the standing waves.  Is special relativity going to
prevail for these standing waves?

As we pointed out in this section, Wigner's formulation of the
$O(3)$-like little group was a very important step.  We are extending
this concept to standing waves.  The development of the quark model
for hadrons was another important step toward understanding Einstein's
covariance~\cite{gell64}.  The proton is a quantum bound state of quarks.
Since the proton these days can achieve a velocity very close to that
of light, it is a relativistic bound-state in the real world.  While the
proton is like a bound state when it is at rest, it appears as a
collection of partons when it moves with velocity close to that of
light.  As we shall discuss in Sec.~\ref{par}, partons have properties
which appear to be quite different from those of quarks.  Can we
produce a standing wave solution for the proton which can explain both
the quark model and the parton model?  This is the problem defined in
the third row of Table~\ref{einwig}.

\section{Can harmonic oscillators be made covariant?}\label{covham}

As we emphasized in Sec.~\ref{feynm}, Quantum field theory has been
quite successful in terms of
perturbation techniques in quantum electrodynamics.  However, this
formalism is based on the S matrix for scattering problems and useful
only for physical processes where a set of free particles becomes
another set of free particles after interaction.  Quantum field theory
does not address the question of localized probability distributions
and their covariance under Lorentz transformations.
The Schr\"odinger quantum mechanics of the hydrogen atom deals with
localized probability distribution.  Indeed, the localization condition
leads to the discrete energy spectrum.  Here, the uncertainty relation
is stated in terms of the spatial separation between the proton and
the electron.  If we believe in Lorentz covariance, there must also
be a time-separation between the two constituent particles.

\begin{figure}[thb]
\centerline{\includegraphics[scale=0.8]{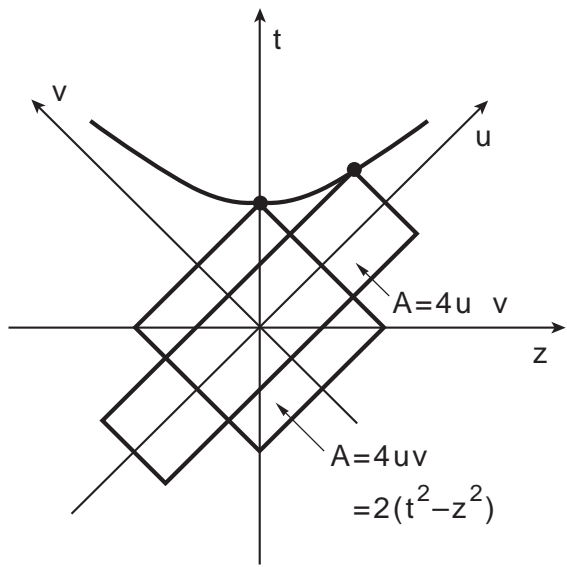}}
\vspace{5mm}
\caption{Lorentz boost in the light-cone coordinate
system.}\label{licone}
\end{figure}

Before 1964~\cite{gell64}, the hydrogen atom was used for
illustrating bound states.  These days, we use hadrons which are
bound states of quarks.  Let us use the simplest hadron consisting of
two quarks bound together with an attractive force, and consider their
space-time positions $x_{a}$ and $x_{b}$, and use the variables
\begin{equation}
X = (x_{a} + x_{b})/2 , \qquad x = (x_{a} - x_{b})/2\sqrt{2} .
\end{equation}
The four-vector $X$ specifies where the hadron is located in space and
time, while the variable $x$ measures the space-time separation
between the quarks.  According to Einstein, this space-time separation
contains a time-like component which actively participates as can be
seen from
\begin{equation}\label{boostm}
\pmatrix{z' \cr t'} = \pmatrix{\cosh \eta & \sinh \eta \cr
\sinh \eta & \cosh \eta } \pmatrix{z \cr t} ,
\end{equation}
when the hadron is boosted along the $z$ direction.
In terms of the light-cone variables defined as~\cite{dir49}
\begin{equation}
u = (z + t)/\sqrt{2} , \qquad v = (z - t)/\sqrt{2} ,
\end{equation}
the boost transformation of Eq.(\ref{boostm}) takes the form
\begin{equation}\label{lorensq}
u' = e^{\eta } u , \qquad v' = e^{-\eta } v .
\end{equation}
The $u$ variable becomes expanded while the $v$ variable becomes
contracted, as is illustrated in Fig.~\ref{licone}.

%------------------------------------------------------------------------
%----------------------------------------------------------------------
Does this time-separation variable exist when the hadron is at rest?
Yes, according to Einstein.  In the present form of quantum mechanics,
we pretend not to know anything about this variable.  Indeed, this
variable belongs to Feynman's rest of the universe.  In this report,
we shall see the role of this time-separation variable in the
decoherence mechanism.

\begin{figure}[thb]
\centerline{\includegraphics[scale=0.8]{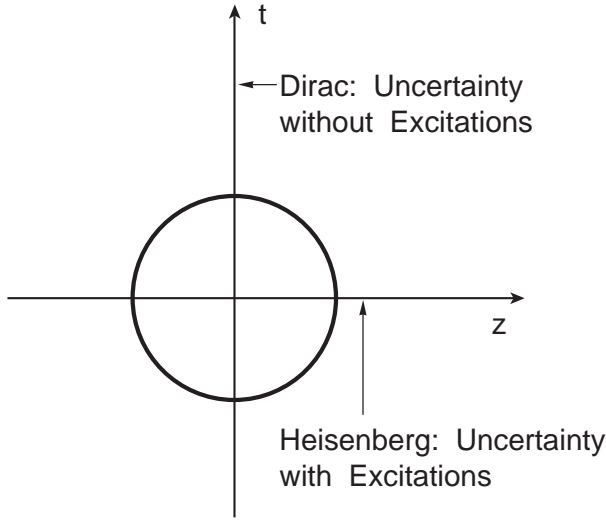}}
\vspace{5mm}
\caption{Space-time picture of quantum mechanics.  There
are quantum excitations along the space-like longitudinal direction, but
there are no excitations along the time-like direction.  The time-energy
relation is a c-number uncertainty relation.}\label{quantum}
\end{figure}

Also in the present form of quantum mechanics, there is an uncertainty
relation between the time and energy variables.  However, there are
no known time-like excitations.  Unlike Heisenberg's
uncertainty relation applicable to position and momentum, the time and
energy separation variables are c-numbers, and we are not allowed to
write down the commutation relation between them.  Indeed, the
time-energy uncertainty relation is a c-number uncertainty
relation~\cite{dir27}, as is illustrated in Fig.~\ref{quantum}

%----------------------------------------------------------------------
%----------------------------------------------------------------------

How does this space-time asymmetry fit into the world of
covariance~\cite{kn73}.  This question was studied in depth by the
present authors in the past.  The answer is that Wigner's $O(3)$-like
little group is not a Lorentz-invariant symmetry, but is a covariant
symmetry~\cite{wig39}.  It has been shown that the time-energy
uncertainty applicable to the time-separation variable fits perfectly
into the $O(3)$-like symmetry of massive relativistic
particles~\cite{knp86}.

The c-number time-energy uncertainty relation allows us to write down
a time distribution function without excitations~\cite{knp86}.
If we use Gaussian forms for both space and time distributions, we
can start with the expression
\begin{equation}\label{ground}
\left({1 \over \pi} \right)^{1/2}
\exp{\left\{-{1 \over 2}\left(z^{2} + t^{2}\right)\right\}}
\end{equation}
for the ground-state wave function.  What do Feynman {\it et al.}
say about this oscillator wave function?

\begin{figure}[thb]
\centerline{\includegraphics[scale=0.4]{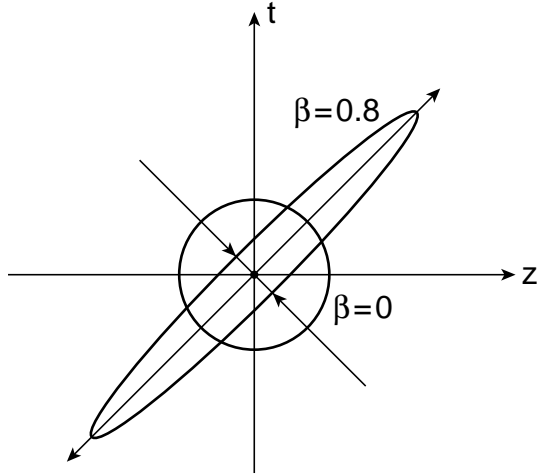}}
\caption{Effect of the Lorentz boost on the space-time
wave function.  The circular space-time distribution in the rest frame
becomes Lorentz-squeezed to become an elliptic
distribution.}\label{ellipse}
\end{figure}

In their classic 1971 paper~\cite{fkr71}, Feynman {\it et al.} start
with the following Lorentz-invariant differential equation.
\begin{equation}\label{osceq}
{1\over 2} \left\{x^{2}_{\mu} -
{\partial^{2} \over \partial x_{\mu }^{2}}
\right\} \psi(x) = \lambda \psi(x) .
\end{equation}
This partial differential equation has many different solutions
depending on the choice of separable variables and boundary conditions.
Feynman {\it et al.} insist on Lorentz-invariant solutions which are
not normalizable.  On the other hand, if we insist on normalization,
the ground-state wave function takes the form of Eq.(\ref{ground}).
It is then possible to construct a representation of the
Poincar\'e group from the solutions of the above differential
equation~\cite{knp86}.  If the system is boosted, the wave function
becomes
\begin{equation}\label{eta}
\psi_{\eta }(z,t) = \left({1 \over \pi }\right)^{1/2}
\exp\left\{-{1\over 2}\left(e^{-2\eta }u^{2} +
e^{2\eta}v^{2}\right)\right\} .
\end{equation}
This wave function becomes Eq.(\ref{ground}) if $\eta$ becomes zero.
The transition from Eq.(\ref{ground}) to Eq.(\ref{eta}) is a
squeeze transformation.  The wave function of Eq.(\ref{ground}) is
distributed within a circular region in the $u v$ plane, and thus
in the $z t$ plane.  On the other hand, the wave function of
Eq.(\ref{eta}) is distributed in an elliptic region with the light-cone
axes as the major and minor axes respectively.  If $\eta$ becomes very
large, the wave function becomes concentrated along one of the
light-cone axes.  Indeed, the form given in Eq.(\ref{eta}) is a
Lorentz-squeezed wave  function.  This squeeze mechanism is
illustrated in Fig.~\ref{ellipse}.

There are many different solutions of the Lorentz invariant differential
equation of Eq.(\ref{osceq}).  The solution given in Eq.(\ref{eta})
is not Lorentz invariant but is covariant.  It is normalizable
in the $t$ variable, as well as in the space-separation variable $z$.
How can we extract probability interpretation from this covariant
wave function?

%----------------------------------------------------------------------

%----------------------------------------------------------------------

\section{Feyman's Parton Picture}\label{par}

It is a widely accepted view that hadrons are quantum bound states
of quarks having localized probability distribution.  As in all
bound-state cases, this localization condition is responsible for
the existence of discrete mass spectra.  The most convincing evidence
for this bound-state picture is the hadronic mass spectra which are
observed in high-energy laboratories~\cite{fkr71,knp86}.

In 1969, Feynman observed that a fast-moving hadron can be regarded
as a collection of many ``partons'' whose properties appear to be
quite different from those of the quarks~\cite{fey69}.  For example,
the number of quarks inside a static proton is three, while the number
of partons in a rapidly moving proton appears to be infinite.  The
question then is how the proton looking like a bound state of quarks
to one observer can appear different to an observer in a different
Lorentz frame?  Feynman made the following systematic observations.

\begin{itemize}

\item[a.]  The picture is valid only for hadrons moving with
  velocity close to that of light.

\item[b.]  The interaction time between the quarks becomes dilated,
   and partons behave as free independent particles.

\item[c.]  The momentum distribution of partons becomes widespread as
   the hadron moves fast.

\item[d.]  The number of partons seems to be infinite or much larger
    than that of quarks.

\end{itemize}

\noindent Because the hadron is believed to be a bound state of two
or three quarks, each of the above phenomena appears as a paradox,
particularly b) and c) together.

In order to resolve this paradox, let us write down the
momentum-energy wave function corresponding to Eq.(\ref{eta}).
If we let the quarks have the four-momenta $p_{a}$ and $p_{b}$, it is
possible to construct two independent four-momentum
variables~\cite{fkr71}
\begin{equation}
P = p_{a} + p_{b} , \qquad q = \sqrt{2}(p_{a} - p_{b}) ,
\end{equation}
where $P$ is the total four-momentum.  It is thus the hadronic
four-momentum.

The variable $q$ measures the four-momentum separation between
the quarks.  Their light-cone variables are
\begin{equation}\label{conju}
q_{u} = (q_{0} - q_{z})/\sqrt{2} ,  \qquad
q_{v} = (q_{0} + q_{z})/\sqrt{2} .
\end{equation}
The resulting momentum-energy wave function is
\begin{equation}\label{phi}
\phi_{\eta }(q_{z},q_{0}) = \left({1 \over \pi }\right)^{1/2}
\exp\left\{-{1\over 2}\left(e^{-2\eta}q_{u}^{2} +
e^{2\eta}q_{v}^{2}\right)\right\} .
\end{equation}
Because we are using here the harmonic oscillator, the mathematical
form of the above momentum-energy wave function is identical to that
of the space-time wave function.  The Lorentz squeeze properties of
these wave functions are also the same.  This aspect of the squeeze
has been exhaustively discussed in the
literature~\cite{knp86,kn77par,kim89}.

When the hadron is at rest with $\eta = 0$, both wave functions
behave like those for the static bound state of quarks.  As $\eta$
increases, the wave functions become continuously squeezed until
they become concentrated along their respective positive
light-cone axes.  Let us look at the z-axis projection of the
space-time wave function.  Indeed, the width of the quark distribution
increases as the hadronic speed approaches that of the speed of
light.  The position of each quark appears widespread to the observer
in the laboratory frame, and the quarks appear like free particles.

%----------------------------------------------------------------------
\begin{figure}%[thb]
\centerline{\includegraphics[scale=0.5]{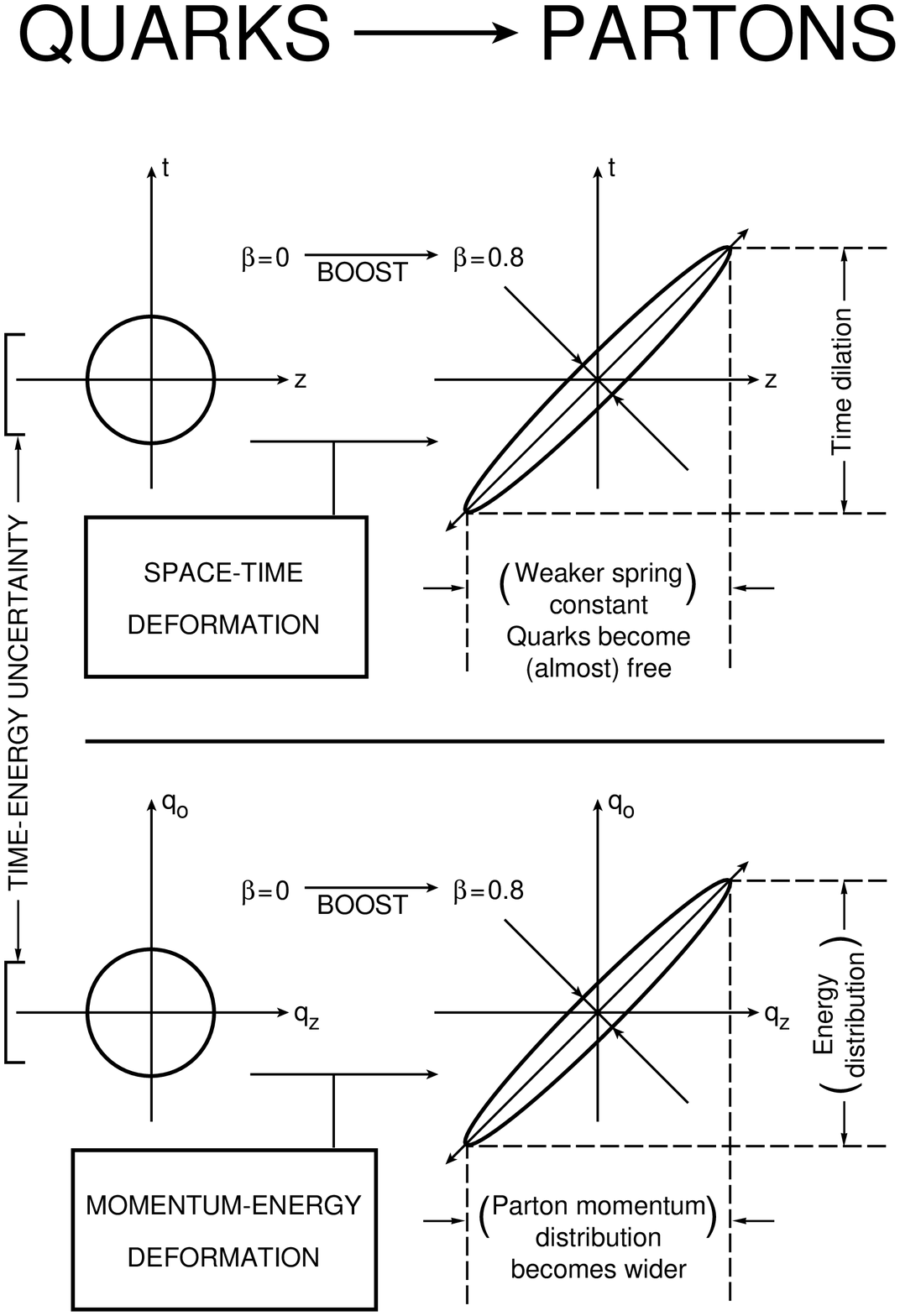}}
\vspace{5mm}
\caption{Lorentz-squeezed space-time and momentum-energy wave
functions.  As the hadron's speed approaches that of light, both
wave functions become concentrated along their respective positive
light-cone axes.  These light-cone concentrations lead to Feynman's
parton picture.}\label{parton}
\end{figure}
%----------------------------------------------------------------------

The momentum-energy wave function is just like the space-time wave
function, as is shown in Fig.~\ref{parton}.  The longitudinal momentum
distribution becomes wide-spread as the hadronic speed approaches the
velocity of light.  This is in contradiction with our expectation from
non-relativistic quantum mechanics that the width of the momentum
distribution is inversely proportional to that of the position wave
function.  Our expectation is that if the quarks are free, they must
have their sharply defined momenta, not a wide-spread distribution.

However, according to our Lorentz-squeezed space-time and
momentum-energy wave functions, the space-time width and the
momentum-energy width increase in the same direction as the hadron
is boosted.  This is of course an effect of Lorentz covariance.
This indeed is the key to the resolution of the quark-parton
paradox~\cite{knp86,kn77par}.

%----------------------------------------------------------------------
\begin{figure}%[thb]
\centerline{\includegraphics[scale=0.6]{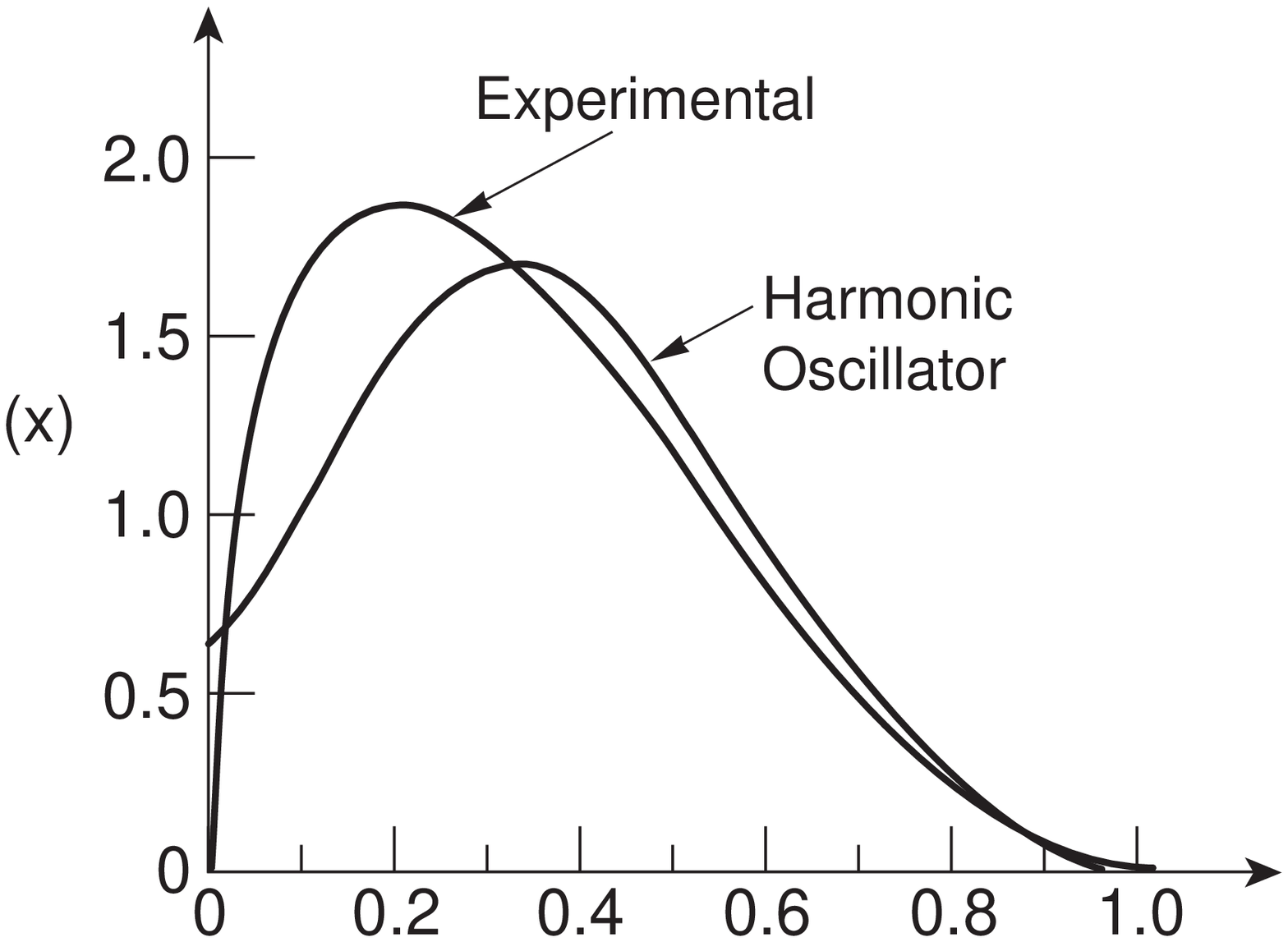}}
\vspace{5mm}
\caption{Parton distribution function.
Theory and experiment.}\label{hussar}
\end{figure}
%----------------------------------------------------------------------

After these qualitative arguments, we are interested in whether
Lorentz-boosted bound-state wave functions in the hadronic rest
frame could lead to parton distribution functions.  If we start with
the ground-state Gaussian wave function for the three-quark wave
function for the proton, the parton distribution function appears
as Gaussian as is indicated in Fig.~\ref{hussar}.  This Gaussian  form
is compared with experimental distribution also in Fig.~\ref{hussar}.

For large $x$ region, the agreement is excellent, but the agreement is
not satisfactory for small values of $x$.  In this region, there is
a complication called the ``sea quarks.''  However, good sea-quark physics
starts from good valence-quark physics.  Figure~\ref{hussar} indicates
that the boosted ground-state wave function provides a good valence-quark
physics.

Feynman's parton picture is one of the most controversial models
proposed in the 20th century.  The original model is valid only in
Lorentz frames where the initial proton moves with infinite momentum.
It is gratifying to note that this model can be produced as a limiting
case of one covariant model which produces the quark model in the
frame where the proton is at rest.  We need Feynman's parton model
to complete the third row of Table~\ref{einwig}.

\section{History, Future, and Strings}\label{future}

In this paper, we are dealing with two different histories.  One
is how to deal with relativistic extended particles starting from
Einstein's special relativity for point particles, as illustrated
in Table~\ref{einwig}.  In so doing,
we had to face another historical problem, namely the problem of
whether scattering problem and bound-state problem can be treated
by the same dynamics, as shown in Table~\ref{scaboun}.  This history
starts from the ancient mystery that comets and planets have
different orbits.

Historically, the unified picture of scattering and bound states
was accomplished by an invention of new dynamics.  As we can see
from Table~\ref{scaboun}, the completion of Newtonian mechanics was
accompanied by a unified view of elliptical and hyperbolic orbits.

At the beginning of the 20th century, discrete energy levels emerged
as one of the most pressing puzzling problems in physics.  Why do
bound states have discrete energy levels while while scattering
states do not.  This question was solved by the wave-particle
duality of quantum mechanics, where bound states satisfy localization
boundary condition.

%----------------------------------------------------------------------
\begin{table}%[thh]
\caption{History of scattering states and bound states.  The history
starts with open and closed orbits of astronomical objects.  Newton
unified the elliptic and hyperbolic orbits with his Newtonian mechanics.
In quantum mechanics with wave-particle duality, running waves and
standing waves tell the difference between bound states and scattering
states.  The remaining problem is whether this quantum picture remains
valid in Einstein's covariant world.}\label{scaboun}
\begin{center}
\vspace{1mm}
\begin{tabular}{rccc}
\hline \\[-3.9mm]
\hline
{}\hspace{7mm} & {}  & {} &  {}\\
{}\hspace{7mm}  & {}& Unified & {} \\
{}\hspace{7mm} & Scattering & Physics & Bound States \\[4mm]\hline
{}\hspace{7mm} & {} & {} & {}\\

Before \hspace{7mm} & {} & {} & {} \\
Newton \hspace{7mm} & Comets & Unknown   & Planets \\[4mm]\hline
{}\hspace{7mm} & {} & {} & {}\\
Newton \hspace{7mm} & Hyperbola & Newton & Ellipse \\[4mm]\hline
{}\hspace{7mm} & {} & {} & {} \\
{}\hspace{7mm} & {} & {} & Quantized \\
Bohr \hspace{7mm} & Unknown & Unknown & Orbits \\[4mm]\hline
{}\hspace{7mm} & {} & {} & {}\\
Quantum \hspace{7mm} & Running  & Particle & Standing \\
Mechanics \hspace{7mm} & Waves & Waves  & Waves \\[4mm]\hline
{}\hspace{7mm} & {} & {} & {}\\
Feynman \hspace{7mm} & Diagrams & Unknown & Oscillators \\[4mm]\hline
{}\hspace{7mm} & {} & {} & {}\\
Future \hspace{7mm} & Running Waves  & One  & Standing Waves \\
Theory \hspace{7mm} & * Fields  & Physics  & * Strings \\[2mm]
\hline
\hline
\end{tabular}
\end{center}
\end{table}
%----------------------------------------------------------------------

In the world of Einstein, the scattering problem is now well understood
in terms of quantum field theory and Feynman diagrams.  In this paper,
we studied whether the covariant harmonic oscillator formalism could
serve as a model for relativistic bound states.  We strengthened our
earlier assertion that that it satisfies every known physical principle
as quantum field theory does~\cite{hknfp81}.  In this report, we
discussed the same problem with the space-time symmetry of standing
waves in the framework of Wigner's little group for massive particles.

We are of course aiming at a unified covariant theory which will take
care of both scattering and bound-state problems.  In the case of the
Schr\"odinger quantum mechanics, we start with one differential
equation, and the difference comes from the boundary condition dictated
by localization of probability distribution.

In the covariant covariant world of quantum mechanics, the story is the
same.  For free particles, the Lorentz-invariant Klein-Gordon equation
is the starting point.  The covariant oscillator formalism starts also
with a Lorentz-invariant differential equation.  The main difference
between running and standing waves is in the boundary conditions, as in
the case of the Schr\"odinger quantum mechanics.  In field theory, we
talk about asymptotic conditions where particles are free particles in
the remote past and remote future in the Lorentz-covariant world, where
causality is preserved.  In the oscillator formalism, we talk about
localization boundary conditions in the Lorentz-covariant world.

Finally, let us make a comment on current activities in string theory.
The ultimate purpose of string theories is to understand space-time
symmetry of particles with internal structures.  As we pointed out
in Sec.~\ref{symm}, this symmetry was worked out by Wigner in his
fundamental paper of 1939.  His work is totally consistent with
Einstein's covariant world.

In addition, in string theories, there are internal vibrations within
particles.  For vibrational problems, we are not aware of any simpler
model than harmonic oscillators.  Let us keep in mind that, in both
engineering and science, it is customary to reduce all complicated
vibrational problems into simple harmonic oscillators before making
contacts with the real world.  Therefore, the most urgent problem in
string theories is to reduce the problem to a soluble model, namely
the covariant harmonic oscillator formalism presented in this report.

This paper indeed provides a place for string theory in the roadmap of
relativity and quantum mechanics, as illustrated in Table~\ref{scaboun}.

\end{document}